\documentclass[osajnl,twocolumn,showpacs,10pt]{revtex4-1} 

\usepackage{amsmath,amssymb,graphicx}

\usepackage{color}

\graphicspath{{pict/}{}}

\newcommand\pictc[5]{\begin{figure}
                   \centerline{
                   \includegraphics[width=#1\columnwidth,height=0.8\textheight,keepaspectratio]{#3}}
               \protect\caption{\protect\label{fig:#4} #5}
                \end{figure}            }

\newcommand\pict[4][1]{\pictc{#1}{!tb}{#2}{#3}{#4}}

\newcommand\rpict[1]{\ref{fig:#1}}

\newcommand\leqt[1]{\protect\label{eq:#1}}
\newcommand\reqtn[1]{\ref{eq:#1}}
\newcommand\reqt[1]{(\reqtn{#1})}

\begin{document}

\title{Nonlinear coupled-mode theory for periodic plasmonic waveguides and metamaterials with loss and gain}

\author{Andrey A. Sukhorukov}
\email[]{Andrey.Sukhorukov@anu.edu.au}

\author{Alexander S. Solntsev}
\author{Sergey~S.~Kruk}
\author{Dragomir N. Neshev}
\author{Yuri S. Kivshar}

\affiliation{Nonlinear Physics Centre and Centre for Ultrahigh-bandwidth Devices for Optical Systems (CUDOS), Research School of Physics and Engineering, Australian National University, Canberra ACT 0200, Australia}

\begin{abstract}
We derive general coupled-mode equations describing the nonlinear interaction of electromagnetic modes in media with
loss and gain. Our approach is rigorously based on the Lorentz reciprocity theorem, and it can be applied to a broad
range of metal-dielectric photonic structures, including plasmonic waveguides and metamaterials. We verify
that our general results agree with the previous analysis of particular cases, and predict novel effects on
self- and cross-phase modulation in multi-layer nonlinear fishnet metamaterials.
\end{abstract}


\maketitle 

Artificial periodic structures such as photonic crystals and metamaterials offer unique possibilities to control both linear light propagation and nonlinear optical interactions~\cite{Denz:2009:NonlinearitiesPeriodic}. In particular, new regimes of nonlinear wave mixing between forward and backward waves can be efficiently realized in negative-index metamaterials~\cite{Shadrivov:2006-529:JOSB, Maimistov:2007-265:RAR, Popov:2010-2999:PB}.
Theoretical analysis of wave propagation in metamaterials is usually based on complimentary numerical and analytical methods. Numerical finite-difference modeling can provide detailed information about the linear eigenmodes and their dispersion, and then the evolution of arbitrary wavefronts can be exactly described in the linear regime according to the superposition principle. In the nonlinear regime, superposition principle does not hold, and approximate analytical methods become essential to describe nonlinear wave mixing under a broad range of varying input conditions, whereas direct numerical simulations can be used to verify only a limited subset of cases.

A powerful analytical method for simulating nonlinear wave propagation in periodic waveguides and metamaterials is based on the derivation of coupled-mode equations for the amplitudes of linear eigenmodes, where the amplitudes can change due to optical nonlinearity. This approach is well established for conventional dielectric waveguides including solid-core optical fibers~\cite{Snyder:1983:OpticalWaveguide}. It was more recently extended to photonic-crystal waveguides~\cite{Michaelis:2003-65601:PRE, Iliew:2008-115124:PRB, Panoiu:2010-257:ISQE}, microstructured fibers~\cite{Shahraam:2009-2298:OE}, and magnetoelectric metamaterials~\cite{Rose:2012-33816:PRA}. In these studies, it is commonly assumed that losses are absent, or there is a small loss which can be considered as a perturbation. It was demonstrated that the effects of loss in arbitrary periodic waveguides can be properly described in the linear regime through the scattering matrix formulation~\cite{Lecamp:2007-11042:OE}.
Whereas expressions for the nonlinear response terms obtained for lossless structures are commonly used empirically for various metal-dielectric structures in presence of absorption~\cite{Feigenbaum:2007-674:OL, Davoyan:2009-21732:OE, Popov:2010-2999:PB},
a rigorous derivation for planar and circular plasmonic waveguides~\cite{Marini:2010-33850:PRA, Marini:2011-63839:PRA} has shown that effective nonlinear response can become complex due to the presence of loss. Effects of linear loss were also shown to be important for frequency conversion in plasmonic waveguides~\cite{Ruan:2009-13502:OE}.

In this Letter, we formulate a rigorous systematic procedure based on the Lorentz reciprocity theorem for obtaining coupled-mode equations, which is suitable for periodic photonic structures and metamaterials made of non-magnetic dielectrics and metals with arbitrary spatially distributed loss, gain, and nonlinear characteristics.
We show that our approach agrees with previous results based on a different method for planar plasmonic structures, and present a novel derivation of nonlinear self- and cross-phase modulation coefficients in a multi-layered fishnet metamaterial.

To derive the coupled-mode equations, we employ the Lorentz reciprocity theorem~\cite{Snyder:1983:OpticalWaveguide}. Conjugated form of the theorem is applicable to lossless structures, and it was previously used to derive modal amplitude equations for linear and nonlinear dielectric photonic-crystal waveguides~\cite{Michaelis:2003-65601:PRE, Iliew:2008-115124:PRB, Panoiu:2010-257:ISQE}. On the other hand, unconjugated form of the Lorentz reciprocity theorem is valid in presence of absorption, and it was used to determine the
modal amplitudes accounting for the effects of linear loss and scattering in periodic waveguides~\cite{Lecamp:2007-11042:OE} and describe nonlinear frequency conversion in plasmonic waveguides~\cite{Ruan:2009-13502:OE}.
In our analysis, we generalize the previous approaches and perform the derivation of nonlinear coupled-mode equations for periodic structures with loss and gain.

We consider periodic structures made of non-magnetic materials ($\mu=1$) with linear susceptibility $\varepsilon(x,y,z) \equiv \varepsilon(x,y,z+d)$, where $(x,y,z)$ are the spatial coordinates and $d$ is the period. In the linear regime the electro-magnetic field at frequency $\omega$ can be represented as a superposition of Bloch eigenmodes, which complex electric and magnetic fields are ${\bf \{E_m,H_m\}}(x,y,z) = {\bf \{e_m,h_m\}}(x,y,z) \exp(i \beta_m z)$. Here $m$ is the number of Bloch mode, $\beta_m$ are the Bloch wavenumbers, and ${\bf \{e_m,h_m\}}$ are the Bloch profiles which are periodic along the $z$ direction, i.e. ${\bf \{e_m,h_m\}}(x,y,z+d) \equiv {\bf \{e_m,h_m\}}(x,y,z)$.
To simplify the notation, we choose the pairs of numbers $(+m, -m)$ to denote counter-propagating waves, such that $\beta_{+m} \equiv - \beta_{-m}$.

We formulate the unconjugated form of the Lorentz reciprocity theorem~\cite{Snyder:1983:OpticalWaveguide, Lecamp:2007-11042:OE} in a way suitable for periodic structures in analogy to the conjugated form derivation for photonic crystals in Ref.~\cite{Michaelis:2003-65601:PRE}:
\begin{equation} \leqt{Lorentz}
   \oint_S {\bf F}_q \cdot d {\bf A} = - i \omega \int_V {\bf E}_q \cdot {\bf P}_{\rm nl} d V ,
\end{equation}
where $q$ is an index of Bloch mode in linear medium, ${\bf P}_{\rm nl}$ describes the medium polarization due to nonlinearity,
${\bf F}_q = {\bf E} \times {\bf H}_q - {\bf E}_q \times {\bf H}$, $\{{\bf E}, {\bf H}\}$ is the electro-magnetic field under the presence of ${\bf P}_{\rm nl}$, $S$ is the surface surrounding volume $V$. We define the integration volume to include a single unit cell in the propagation direction, $(z-d/2,z+d/2)$. In the transverse directions, we also choose a single unit cell if the structure is periodic in $(x,y)$ plane, or extend the integration to infinity if the waveguide supports transversely localized modes.

We formulate the bi-orthogonality property of unperturbed modes by analyzing the case when ${\bf P}_{\rm nl} \equiv 0$ and $\{{\bf E}, {\bf H}\} = \{{\bf E}_m, {\bf H}_m\}$. Then, using the periodicity conditions, we obtain that $\int dx dy [{\bf e}_m \times {\bf h}_q - {\bf e}_q \times {\bf h}_m]_z = s_m = - s_{-q}$ for $q=-m$, and the integral vanishes for $q \ne -m$. Importantly, the value of $s_m$ does not depend on $z$, and it has the meaning of adjoint flux~\cite{Chen:2010-53825:PRA}.

Next, we seek an approximate solution in the nonlinear regime as a superposition of linear modes with amplitudes $a_m(z)$,
\begin{equation} \leqt{modes}
  {\bf \{E,H\}}(x,y,z) = \sum_m a_m(z) {\bf \{e,h\}}_m(x,y,z) .
\end{equation}
We assume that the nonlinearity is relatively weak, which is expressed mathematically as the condition that ${\tilde a}_m(z) = a_m(z) \exp(- i \beta_m z)$ change slowly over each structure period  (in the absence of nonlinearity, ${\tilde a}_m$ would be constant). Then,
we perform Taylor expansion of the field amplitude as $a_m(z + \xi) = \exp[i \beta_m (z+\xi)] {\tilde a}_m(z + \xi) \simeq \exp[i \beta_m (z+\xi)] [ {\tilde a}_m(z) + \xi d {\tilde a}_m(z) / d z] = \exp(i \beta_m \xi) \{ a_m(z) + \xi [ d a_m(z) / d z - i \beta_m a_m(z)] \}$. We substitute Eq.~\reqt{modes} into Eq.~\reqt{Lorentz}, apply the bi-orthogonality relations, use the amplitude expansion, and finally obtain the coupled-mode equations,
\begin{equation} \leqt{CMT}
\begin{split}
  \frac{d a_m}{d z} - i \beta_m a_m
   = & \frac{-i \omega}{s_m d} \int\int dx dy \int_{-d/2}^{d/2} d\xi e^{-i\beta_m \xi} \\
   &
     {\bf e}_{-m}(x,y,z+\xi) \cdot {\bf\tilde P}_{\rm nl}(x,y,z,\xi) ,
\end{split}
\end{equation}
where we neglect the higher-order term $[\xi d {\tilde a}_m(z) / d z]$ under the integral expression on the right-hand side, which is an established approach~\cite{Iliew:2008-115124:PRB}.
The nonlinear polarization under the integral is defined as
${\bf\tilde P}_{\rm nl}(x,y,z,\xi) = {\bf P}_{\rm nl}\left[ {\bf\tilde E}(x,y,z,\xi), {\bf\tilde H}(x,y,z,\xi), x, y, z+\xi \right]$, where ${\bf P}_{\rm nl}$ is dependent on the material properties and the field amplitudes at the particular spatial location, and the fields within a period are defined at the lowest order of Taylor expansion as ${\bf \{\tilde E,\tilde H\}}(x,y,z,\xi) = \sum_q a_q(z) \exp(i \beta_q \xi) {\bf \{e_q,h_q\}}(x,y,z+\xi)$.

In symmetric structures, where $\varepsilon(x,y,z) \equiv \varepsilon(x,y,-z)$, there is a simple relation between the
fields of counter-propagating modes~\cite{Lecamp:2007-11042:OE}:
  $e_{-m}^{(x,y)}(x,y,z) = e_{m}^{(x,y)}(x,y,-z)$, $e_{-m}^{(z)}(x,y,z) = -e_{m}^{(z)}(x,y,-z)$, $h_{-m}^{(x,y)}(x,y,z) = - h_{m}^{(x,y)}(x,y,-z)$, and $h_{-m}^{(z)}(x,y,z) = h_{m}^{(z)}(x,y,-z)$.
Therefore in such structures, it is not necessary to separately determine the profiles of backward-propagating modes to calculate the overlap integrals in Eq.~\reqt{CMT}. We checked that Eqs.~\reqt{CMT} coincide with equations obtained by Ruan {\em et al.} using unconjugated reciprocity relation for frequency conversion in plasmonic waveguides~\cite{Ruan:2009-13502:OE}, where this symmetry is satisfied.
We also note that in lossless structures $e_{-m}(x,y,z) = e_{m}^\ast(x,y,z)$, and in such case Eq.~\reqt{CMT} reduces to the previously derived equations based on conjugated reciprocity relation~\cite{Iliew:2008-115124:PRB}.

We first demonstrate the application of our general approach to a simple structure, considering a surface plasmon polariton (SPP) at a planar metal-dielectric interface. This allows us to perform a comparison of our theory with the previous results~\cite{Marini:2010-33850:PRA} obtained through direct perturbation analysis. In these studies, careful derivation was carried out taking into account the metal-dielectric boundary conditions, where some field derivatives abruptly change sign. It is therefore an important testcase to validate our approach.

We analyze the SPP at a metal-dielectric interface positioned at $x=0$, which is homogeneous in the propagation ($z$) and transverse ($y$) directions. Accordingly, there is no periodicity and formally $d \rightarrow 0$. Following Marini and Skryabin~\cite{Marini:2010-33850:PRA}, we consider a dielectric response with linear gain and Kerr nonlinear susceptibility $\varepsilon_{\rm nl} = \chi |{\bf E}|^2$. The linear field profile of propagating SPP in the dielectric ($x>0$) is:
$e^{(x)} = i \beta \exp(-\rho_d x) / \rho_d$, $e^{(z)} = \exp(-\rho_d x)$, $h^{(y)} = - i \omega \epsilon_d \exp(-\rho_d x) / \rho_d$,
and in the metal ($x<0$): $e^{(x)} = - i \beta \exp(\rho_m x) / \rho_m$, $e^{(z)} = \exp(\rho_m x)$, $h^{(y)} = i \omega \epsilon_m \exp(\rho_m x) / \rho_m$.
Here $\epsilon_d$ and  $\epsilon_m$ are the linear permittivities of dielectric and metal, respectively, $\beta = \omega c^{-1} [ \epsilon_d \epsilon_m / (\epsilon_d + \epsilon_m) ]^{1/2}$, $\rho_{d,m}^2 = \beta^2 - \omega^2 c^{-2} \epsilon_{d,m}$, and $c$ is the speed of light in vacuum.
Then, from Eq.~\reqt{CMT} we obtain the following equation for the amplitude of SPP, $d a / d z - i \beta a = i \gamma_{\rm SPM} |a|^2 a$. The effective self-phase-modulation (SPM) coefficient is found as
$\gamma_{\rm SPM} = -(\omega/s) \chi \int dx (|e^{(x)}|^2 + |e^{(z)}|^2)[(e^{(x)})^2 - (e^{(z)})^2]$, where $s = - 2 \int dx e^{(x)} h^{(y)}$.
After integration, we obtain an expression for the nonlinear term, which is equivalent to the formula for $\gamma$ in Eq.~(28) of Ref.~\cite{Marini:2010-33850:PRA}.
This demonstrates that the coupled-mode Eqs.~\reqt{CMT} based on Lorentz reciprocity relations provide a description consistent with multi-step perturbation series expansion, with the advantage of being directly applicable for complex waveguide geometries.

Our approach can be rigorously applied to three-dimensional metamaterials. The linear profiles of Bloch modes in such structures can be readily calculated numerically 
through finite-difference time-domain simulations (FDTD) combined with high-resolution spectral analysis taking into account the Bloch wave symmetry properties~\cite{Andryieuski:2012-35127:PRB}. Then the coupled-mode equations are immediately obtained through a calculation of the overlap integrals according to Eq.~\reqt{CMT}.
Whereas a metal nonlinearity can play an important role, in order to illustrate the effect of loss on the effective nonlinear response, we analyze the nonlinear polarization due to a nonlinear dielectric embedded in the structure with metal elements. We choose the nonlinear dielectric susceptibility to be purely real, according to a conventional form~\cite{*[{}] [{, Sec. 4.2.}] Boyd:2008:NonlinearOptics}:
${\bf P}_{\rm nl} = \chi_A ({\bf E} \cdot {\bf E}^\ast ) {\bf E} + (\chi_B / 2) ({\bf E} \cdot {\bf E} ) {\bf E}^\ast$. Considering the interaction of a mode with the same mode propagating in opposite direction, we obtain coupled-mode equations while neglecting quickly oscillating terms:
\begin{equation} \leqt{CM-fishnet}
 \begin{split}
   {d a_+}/{d z} & - i \beta a_+ = i (\gamma_{\rm SPM} |a_+|^2 + \gamma_{\rm XPM} |a_-|^2) a_+ , \\
   {d a_-}/{d z} & + i \beta a_- = - i (\gamma_{\rm SPM} |a_-|^2 + \gamma_{\rm XPM} |a_+|^2) a_- ,
 \end{split}
\end{equation}
where $a_+$ and $a_-$ are the amplitudes of waves propagating in forward (+z) and backward (-z) directions, respectively. The nonlinear self-phase modulation (SPM) and cross-phase modulation (XPM) coefficients are:
\begin{eqnarray}
  &&\gamma_{\rm SPM} = \frac{-\omega}{s_+ d} \int\int dx dy \int_{-d/2}^{d/2} d z\,
    {\bf e}_{-}(x,y,z) \nonumber \\
    &&\, \cdot \left[ \chi_A ({\bf e}_+ \cdot {\bf e}_+^\ast ) {\bf e}_+ + \chi_B ({\bf e}_+ \cdot {\bf e}_+ ) {\bf e}_+^\ast / 2 \right] ,
       \leqt{gamma-fishnet} \\
  &&\gamma_{\rm XPM} = \frac{-\omega}{s_+ d} \int\int dx dy \int_{-d/2}^{d/2} d z\,
    {\bf e}_{-}(x,y,z) \nonumber \\
    &&\cdot \left[ \chi_A ({\bf e}_- \cdot {\bf e}_-^\ast ) {\bf e}_+
            + \chi_A ({\bf e}_+ \cdot {\bf e}_-^\ast ) {\bf e}_-
            + \chi_B ({\bf e}_+ \cdot {\bf e}_- ) {\bf e}_-^\ast \right]. \nonumber
\end{eqnarray}
We note that in the absence of losses, ${\bf e}_-(x,y,z) = {\bf e}_+^\ast(x,y,z)$, and all terms under the integrals become real. However when losses (or gain) are present, such relations no longer hold, and the effective nonlinear coefficients can become complex even when the nonlinear susceptibility parameters $\chi_A$ and $\chi_B$ are real-valued.

As an example, we consider a fishnet metamaterial filled with chalcogenide glass as shown in Fig.~\rpict{fishnet}(a). We note that a multilayer fishnet is the only experimentally realized 3D metamaterial structure with negative refractive index at optical frequencies~\cite{Valentine:2008-376:NAT}. In the numerical analysis we assume that the refractive index of the chalcogenide glass is 2.35~\cite{Todorov:2011-305401:JPD}, while the dispersion of the refractive index of silver~\cite{Johnson:1972-4370:PRB} and silicon~\cite{Thutupalli:1977-467:JPC} are fully taken into account. To simplify the calculations we neglect the nonlinear effects in Silicon, because the electric field appears to be mostly concentrated in the holes filled with chalcogenide glass, which has strong Kerr optical nonlinearity. We carry out FDTD simulations with CST Microwave Studio, using plane-wave excitation with electric field linearly polarized in $y$ direction.
We determine the effective linear refractive index by using the Bloch mode extraction approach~\cite{Andryieuski:2012-35127:PRB} and inverted Fresnel formulae~\cite{Smith:2002-195104:PRB}. The data obtained by both methods is in good agreement with each other as shown in Fig.~\rpict{fishnet}(b).
The first approach also allows us to determine the profiles of linear Bloch waves. In Fig.~\rpict{field} we show examples of electric field profiles for the Bloch modes at wavelengths corresponding to a negative (at $\lambda = 1550\,$nm) and two positive (at $1250\,$nm and $1047\,$nm) refractive indices.
The $y$-components of the electric field are similar for all three wavelengths, whereas $z$-components are substantially different; $x$-components are negligible for $y$-polarized excitation.

\pict{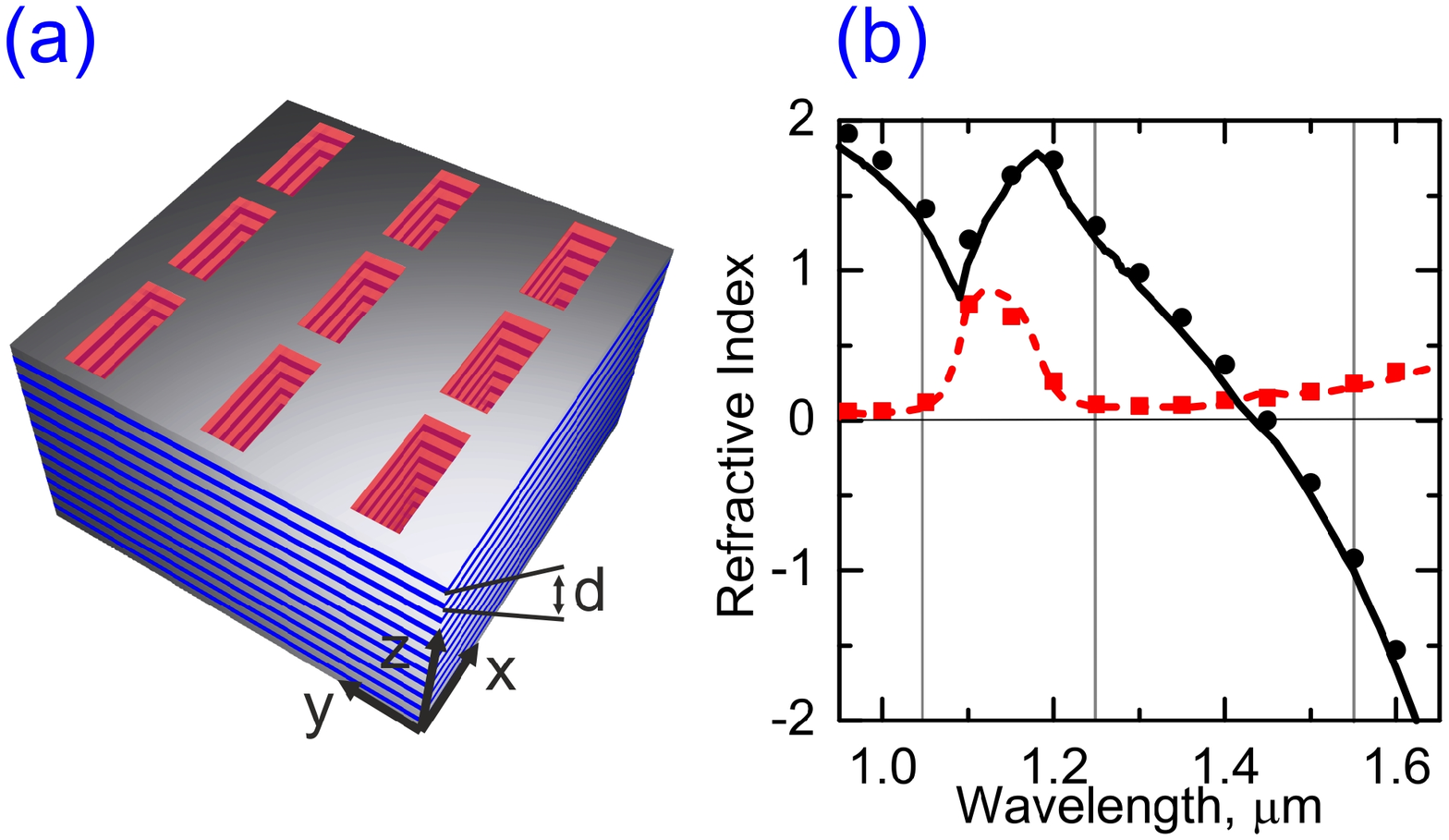}{fishnet}{
(a)~Scheme of a multi-layer fishnet metamaterial consisting of 45~nm thick silver layers separated by 30~nm thick silicon layers ($d = 75\,$nm). The $110 \times 190$~nm holes are filled with chalcogenide glass, and positioned with periodicity of 270~nm in both lateral directions $(x,y)$.
(b)~Effective linear refractive index: real (black) and imaginary (red) parts. Lines are calculated using Bloch modes formalism~\cite{Andryieuski:2012-35127:PRB}, dots~--- by inverted Fresnel formula~\cite{Smith:2002-195104:PRB}.
}

\pict[1]{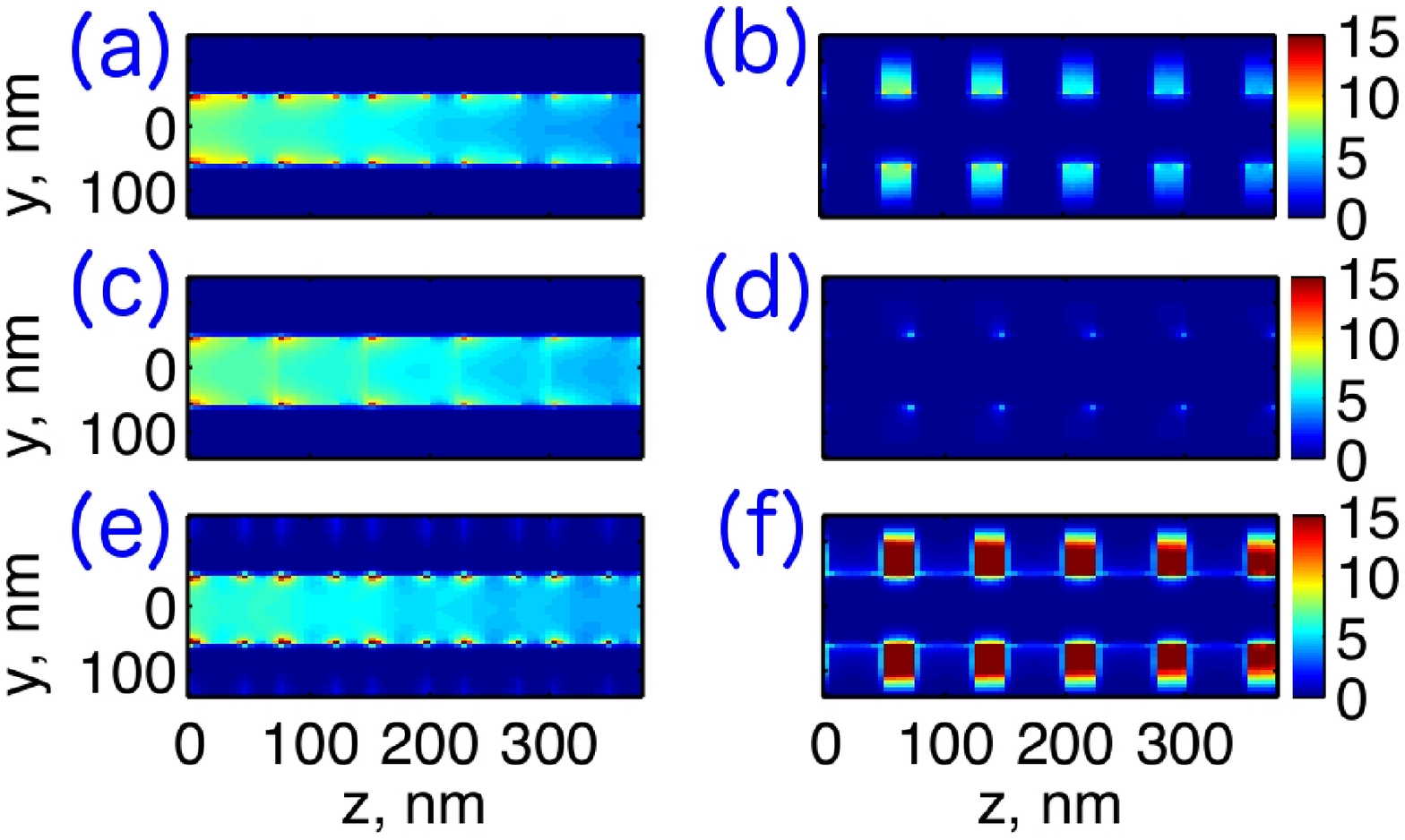}{field}{
Absolute values of the electric fields in a $(y,z)$ cross-section aligned with a hole center for linear Bloch modes at different optical wavelengths: (a,b)~1550~nm, (c,d)~1250~nm, (e,f)~1047~nm. Shown are the (a,c,e)~$y$-components and (b,d,f)~$z$-components.
The fields are normalized to a unity energy flow across one period in the $(x,y)$ cross-section.
}

For chalcogenide glass, the nonlinearity is due to a nonresonant electronic response, and therefore $\chi_A = \chi_B \equiv (2/3) \chi$~\cite{*[{}] [{, Sec. 4.2.}] Boyd:2008:NonlinearOptics}.
Additionally, our structure has the symmetry $z\rightarrow -z$, and using the corresponding field symmetry formulated above it can be shown that $({\bf e}_- \cdot {\bf e}_-^\ast )|_z = ({\bf e}_+ \cdot {\bf e}_+^\ast )|_{-z}$, $({\bf e}_- \cdot {\bf e}_- )|_z = ({\bf e}_+ \cdot {\bf e}_+ )|_{-z}$, $({\bf e}_+ \cdot {\bf e}_- )|_z = ({\bf e}_+ \cdot {\bf e}_- )|_{-z}$, $({\bf e}_+ \cdot {\bf e}_-^\ast )|_z = ({\bf e}_+^\ast \cdot {\bf e}_- )|_{-z}$, and subsequently
$\gamma_{\rm XPM} = 2 \gamma_{\rm SPM}$.
We calculate the effective SPM nonlinear coefficients for the fishnet structure, and compare them with the values for a bulk chalcogenide glass under the condition of the same optical power flow. The results are summarized in Table~\ref{tab:table} for different wavelengths, corresponding to the regimes of negative and positive effective linear refractive index values. We see that in fishnet structure, the effective nonlinear coefficient is enhanced, due to a strong field confinement in the dielectric. On the other hand, the coefficient becomes complex. This means that nonlinearity reduces or increases losses, depending on the sign of ${\rm Im}(\gamma)$. Intuitively, it can be understood because the nonlinearity effectively shifts the dispersion curves~\cite{Soljacic:2004-211:NMAT} and changes the absorption due to the frequency dependant losses.

\begin{table}[t]
\begin{center}
    \begin{tabular}{| l | l | l | l |}
    \hline
    $\lambda$, ${\rm \mu m}$ & $n_{\rm eff}$ & $\beta$, 10$^6$m$^{-1}$
& $\gamma_{\rm SPM} / \gamma_{\rm bulk}$ \\ \hline
    $1.550$ & $-1.02+ i 0.22$ & $-4.13 + i 0.88$ & $35.66 - 1.64 i$ \\ \hline
    $1.250$ & $1.24 + i 0.14$ & $6.22 + i 0.68$ & $15.84 - 0.32 i$ \\ \hline
    $1.047$ & $1.30 + i 0.10$ & $7.80 + i 0.57$ & $7.74 + 0.24 i$ \\ \hline

    \end{tabular}
\end{center}
\caption{Calculated parameters for the fishnet metamaterial at three
different wavelengths $\lambda$: effective refractive index $n_{\rm
eff}$, propagation constant $\beta$, SPM coefficient normalized to its
bulk value.}
    \label{tab:table}
\end{table}

In conclusion, we have developed a systematic approach for the rigorous derivation of the coupled-mode equations in nonlinear periodic structures with loss and gain. This formulation is based on the Lorentz reciprocity theorem, and it allows us to accurately determine the effective nonlinear coefficients.
The method is computationally efficient, and it can be applied to the analysis of a variety of plasmonic waveguides and three-dimensional metamaterials~\cite{Valentine:2008-376:NAT}. The derived coupled-mode equations can be used to describe multi-wavelength interactions at different frequencies, such as in the process of harmonic generation and four-wave mixing~\cite{Suchowski:2013-QM1A.2:ProcCLEO}.

We acknowledge support from the Australian Research Council including Future Fellowship FT100100160 and Discovery Project DP130100135, and Australian NCI National Facility.

\end{document}